\documentclass[prd,twocolumn,superscriptaddress,preprintnumbers,nofootinbib]{revtex4}
\usepackage{graphicx}
\usepackage{epsfig}
\usepackage{bm}
\usepackage{latexsym,amssymb,amsmath,amsfonts,amssymb,wasysym,float}
\usepackage{scrextend}
\usepackage{mathrsfs}
\usepackage{color}

\usepackage{enumitem}

\newcommand{\postscript}[2]{\setlength{\epsfxsize}{#2\hsize}
   \centerline{\epsfbox{#1}}}

\usepackage[usenames,dvipsnames]{xcolor}
\definecolor{orange}{cmyk}{0,0.5,1,0}
\definecolor{rossoCP3}{cmyk}{0,.88,.77,.40}
\definecolor{graa}{rgb}{0.8,0.8,0.8}
\definecolor{blaa}{rgb}{0.2,0.2,0.6}
\usepackage{xspace} 
\newcommand{\CLASS}{\texttt{CLASS}\xspace}
\newcommand{\MP}{\texttt{MontePython}\xspace}

\begin{document}
\preprint{MPP-2023-288}
\preprint{LMU-ASC 39/23}

\title{\color{rossoCP3}
Anti-de Sitter $\bm{\to}$ de Sitter transition driven by   
  Casimir forces and mitigating tensions in cosmological parameters
}

\author{\bf Luis A. Anchordoqui}

\affiliation{Department of Physics and Astronomy,  Lehman College, City University of
  New York, NY 10468, USA
}

\affiliation{Department of Physics,
 Graduate Center, City University
  of New York,  NY 10016, USA
}

\affiliation{Department of Astrophysics,
 American Museum of Natural History, NY
 10024, USA
}

\author{\bf Ignatios Antoniadis}
\affiliation{Laboratoire de Physique Th\'eorique et Hautes \'Energies - LPTHE
Sorbonne Universit\'e, CNRS, 4 Place Jussieu, 75005 Paris, France
}
\affiliation{Center for Cosmology and Particle Physics, Department
  of Physics, New York University, 726 Broadway, New York, NY 10003, USA}

\author{\bf Dieter\nolinebreak~L\"ust}

\affiliation{Max--Planck--Institut f\"ur Physik,  
 Werner--Heisenberg--Institut,
80805 M\"unchen, Germany
}

\affiliation{Arnold Sommerfeld Center for Theoretical Physics, 
Ludwig-Maximilians-Universit\"at M\"unchen,
80333 M\"unchen, Germany
}

\begin{abstract}
  \vskip 2mm \noindent Over the last few years, low- and high-redshift
  observations set off tensions in the measurement of the present-day
  expansion rate $H_0$ and in the determination of the amplitude of the matter clustering
  in the late Universe (parameterized by $S_8$). It was recently noted
  that both these tensions can be resolved if the cosmological constant
  parametrizing the dark energy content switches its sign at a
  critical redshift $z_c \sim 2$. However, the anti-de Sitter (AdS)
  swampland conjecture suggests that the postulated switch in sign of the
  cosmological constant at zero temperature seems unlikely because the
  AdS vacua are an infinite distance appart from de Sitter (dS)
  vacua in moduli space. We provide an explanation for the required AdS
  $\to$ dS crossover transition in the vacuum energy using the Casimir
  forces of fields inhabiting the bulk. We then use entropy arguments to
  claim that any AdS $\to$ dS transition between metastable vacua
  must be accompanied by a reduction of the species scale where gravity
  becomes strong. We provide a few examples supporting this AdS $\to$
  dS uplift conjecture.
\end{abstract}

\maketitle 

It has been almost a century since the expansion of the Universe was
established~\cite{Hubble:1929ig}, but the Hubble constant $H_0$, which measures its rate, 
continues to encounter challenging shortcomings. Actually, the
mismatch between the locally measured value of
$H_0$~\cite{Riess:2021jrx} and the one inferred from
observations of the cosmic microwave background (CMB) on the basis of the $\Lambda$ cold dark matter (CDM)
model~\cite{Planck:2018vyg} has become the mainspring of modern
cosmology, and many extensions of $\Lambda$CDM are
rising to the challenge~\cite{DiValentino:2021izs}. Concomitant with
the $H_0$ tension, there is evidence of a growing tension between the
CMB-preferred value~\cite{Planck:2018vyg} and weak gravitational lensing determination~\cite{KiDS:2020suj} of the weighted
amplitude of matter fluctuations parametrized by $S_8$. It
would be appealing and compelling if both the $H_0$ and $S_8$ tensions
were resolved at once, but as yet none of the proposed extensions of
$\Lambda$CDM have done so to a satisfactory
degree~\cite{Abdalla:2022yfr}.

Very recently, an empirical conjecture
to simultaneously resolve the $H_0$ and $S_8$ tensions has been contemplated~\cite{Akarsu:2023mfb}. The
conjecture, which is rooted on the graduated dark energy
model~\cite{Akarsu:2019hmw,Akarsu:2021fol,Akarsu:2022typ}, postulates
that $\Lambda$ may have switched sign (from negative to positive) at
critical redshift $z_c \sim 2$. The so-called ``$\Lambda_s$CDM model'' then suggests that
the Universe may have recently experienced a phase of rapid transition
from anti-de Sitter (AdS) vacuum to de Sitter (dS) vacuum. However, the AdS swampland conjecture~\cite{Lust:2019zwm} indicates that AdS vacua are an infinite distance
apart from dS vacua in moduli space. In fact, the AdS
  distance conjecture, taken at face value, implies that it is not
  possible to cross the barrier $\Lambda=0$ and to have a transition
  from a negative to a positive cosmological constant. This follows because the distance in the space of metrics in terms of  $\Lambda=0$ is proportional to  $-\log|\Lambda|$,
and this expression diverges as  $\Lambda\rightarrow0$. However, we
note that this no-go theorem is valid at zero temperature, where the
number of light particles is (in general) constant. At finite temperature, particles can decay and hence
the number of light particles can change. In this way the minima of
the potential can be lifted. All in all, the
postulated switch in sign of $\Lambda$ at zero temperature seems, at
first sight, unlikely. In this Letter we investigate whether quantum effects derived from
the Casimir energy of fields inhabiting the bulk could produce the hypothesized
AdS $\to$ dS crossover transition.\footnote{For related work, see~\cite{Ye:2020btb}.}    

Before proceeding, we pause to discuss the benefits of the
$\Lambda_s$CDM model. The model simultaneously resolves the major
cosmological tensions including: $H_0$, $S_8$, and SNe Ia absolute magnitude~\cite{Akarsu:2023mfb}. In addition, it can
accommodate the BAO Lyman-$\alpha$
discrepancy~\cite{Akarsu:2019hmw}.\footnote{The recent investigation
  presented in~\cite{Akarsu:2023mfb} is based on: the {\it Planck} CMB
  data~\cite{Planck:2019nip}, the Pantheon+ SNe
  Ia sample~\cite{Brout:2022vxf}, the data release of
  KiDS-1000~\cite{KiDS:2020suj}, and the (angular) transversal 2D BAO
  data on the shell~\cite{Nunes:2020hzy,deCarvalho:2021azj}, which are
  less model dependent than the 3D BAO data used in previous studies
  of $\Lambda_s$CDM. It is important to stress that the BAO
3D data sample assumes $\Lambda$CDM to determine the distance to the
spherical shell, and hence could potentially introduce a bias when
analyzing beyond $\Lambda$CDM models~\cite{Bernui:2023byc}.} Furthermore, for flat cosmology, $\Omega_m +
\Omega_\Lambda \simeq 1$, from which it follows that the mean energy density in the universe is equal to the critical density, where $\Omega_m$ is the fractional
nonrelativistic matter density and $\Omega_\Lambda$ is the
cosmological-constant energy
density. This implies that $\Omega_m >
1$  when $\Lambda < 0$, in agreement with the otherwise puzzling JWST
observations~\cite{Adil:2023ara}.\footnote{See
  however~\cite{Sabti:2023xwo} and~\cite{Padmanabhan:2023esp}.}

In order to demonstrate the main mechanism of the AdS $\to$ dS crossover transition we will first consider a simple toy model.
This toy model already incorporates some of the main features that we will meet in the cosmological model, which will be discussed next, and it is also compatible with the dark dimension scenario~\cite{Montero:2022prj}.
We will start with an AdS potential, which is quadratic in the radius modulus $R$ of a one-dimensional compactification, and which leads to AdS minimum\footnote{Such AdS potentials are expected to arise in the
context of flux compactifications. More generally, the quadratic form is
also justified around the minimum.}:
\begin{equation}
V_{\rm AdS}(R)={1\over R_0^6}(R-R_0)^2-{1\over R_0^4}\, .
\label{uno}
\end{equation}
This potential possesses a minimum at $R=R_0$ and the potential at its minimum satisfies
\begin{equation}
V_{\rm AdS}(R_0)=-{1\over R_0^4}\, .
\end{equation}
Hence this AdS minimum exhibits scale separation in four dimension, in agreement with the weak form of the AdS swampland conjecture.
Next we consider a positive Casimir energy of the simple form 
\begin{equation}
V_C (R,a) ={a\over R^4}\, ,
\label{V1}
\end{equation}
where $a$ is a so far undetermined constant. Here $V_C(R)$ provides
the potential uplift to dS, as the total potential is given by 
\begin{eqnarray}
V(R,a)&=&V_{\rm AdS}(R)+V_C(R)\nonumber\\
&=&{1\over R_0^6}(R-R_0)^2-{1\over R_0^4}+{a\over R^4}\, .
\end{eqnarray}
This potential possesses a minimum at $R=R_{1}$, for which one can show that it  always satisfies $R_{1}>R_0$.
So the radius gets increased due to the uplift, and hence the
associated species
scale~\cite{Dvali:2007hz,Dvali:2007wp,vandeHeisteeg:2022btw,Cribiori:2022nke,vandeHeisteeg:2023dlw}
decreases (see also the discussion below). The species
  entropy was introduced in~\cite{Cribiori:2023ffn}, and it is defined in four space-time
  dimensions as $S_{\rm sp}=M_*^{-2}$. Following the law of thermodynamics, it is also argued in~\cite{Cribiori:2023ffn}
that any dynamical motion in moduli space should always follow a path,
for which $S_{\rm sp}$ does not decrease. With $M_*\simeq R^{-1/3}$, it follows that species thermodynamics 
allows only for transition processes where the radius of the extra dimension increases (or stays constant) due to the uplift.

The value and in particular the sign of the the uplifted vacuum energy $V(R_{1},a)$ now depends on the value of the constant $a$. Actually, one can show that there is  critical value $a^{c}=5^5/3^6\simeq 4.28$ and an associated
critical radius $R_{1}^c=5R_0/3$, where the potential exactly vanishes:
\begin{equation}
V(R_{1}^c,a^c)=0\,.
\end{equation}
For $0<a< a^c$ and $R_0<R_1<R_1^c$ the vacuum is negative: $V(R_{1}^c,a)<0$; for $a>a^c$ and $R_1>R_1^c$ the vacuum becomes positive, i.e. $V(R_{1}^c,a)>0$.
For example, for $a=16$ it follows that the minimum is at $R_1=2R_0$,
and the corresponding positive vacuum energy takes the value
\begin{equation}
V(R_{1}=2R_0,a=16)={1\over R_0^4} \, .
\end{equation}
Note that for $a - a^c \to 0^+$, the minimum $R_1$ is near $R_0$,
consistently with the quadratic approximation given in (\ref{uno}). Thus, the parametric dependence between the vacuum energy and the radius
at its minimum is the same as before the uplift, in accordance with
the dark dimension scenario~\cite{Montero:2022prj}. A point worth
noting at this juncture is that the relation $|V| = R^{-4}$ is at the heart of vacuum energy calculations in string theory, see~\cite{Anchordoqui:2023laz}
for a recent discussion. 

In summary  we have demonstrated that by varying the parameter $a$ of
the Casimir energy, we could model  
a phase transition from negative vacuum energy to positive vacuum
energy, i.e. from AdS to dS.\footnote{Alternative uplift scenarios for
  dS were originally suggested by Kachru, Kallosh, Linde, and Trived~\cite{Kachru:2003aw}.} In the early universe $a$ becomes a function of the temperature, $a=a(T)$, and at a certain critical temperature $T_c$,
there could be a phase transition from from AdS to dS, like we have
described above. Alternatively, for a compact space without cosmic expansion, the dynamics of
$a$ could be driven by the particle decay widths of the spectrum in
the deep infrared region. 

Now, we turn to construct a cosmological model that can simultaneously
resolve the $H_0$ and
$S_8$ tensions; namely 
we compactify a 5-dimensional (5D) model of Einstein
gravity equipped with a next-to-minimal hidden sector. The model is
based on the idea that a compact dimension may have undergone a uniform
rapid expansion, together with the 3D non-compact
space, via regular exponential inflation described by a 5D de Sitter (or approximate) solution of Einstein
equations~\cite{Anchordoqui:2022svl}. The size of the compact space has
grown from the fundamental length ${\cal O} (M_*^{-1})$ to the micron
size, so that at the end of inflation the emergent 4D strength
of gravity became much weaker, as it is measured today,
\begin{equation}
  M_p^2 = 2 \pi R_\perp M_* \ ,
\end{equation}
where $M_p$ is the reduced Planck mass, $R_\perp$ the radius of the
compact space at the end of inflation, and $M_*$ the 5D Planck scale
(or species scale where gravity becomes
strong~\cite{Dvali:2007hz,Dvali:2007wp,vandeHeisteeg:2022btw,Cribiori:2022nke,vandeHeisteeg:2023dlw}). As
a matter of fact, $M_*$ is not a fundamental scale but only the
energy/length region where gravity becomes strong and requires an UV
completion, such as string theory. The string scale is the fundamental
parameter which does not change, but 
as we discuss below, the species scale can vary depending the particle content and interaction at a given energy. In our case the variation is minimal and within the energy region where gravity becomes strong.

In the spirit of~\cite{Schwarz:2024tet}, we model the dark dimension as a space with two
boundaries: an interval with end-of-the-world 9-branes attached at each
end. After a suitable compactification of six dimensions the space 
boundaries can be approximated by 3-branes, and one of them can be
identified with the Standard Model 3-brane which portrays our
non-compact 4D world, at least up to energies ${\cal O} (10~{\rm TeV})$ and perhaps much further. The line interval along the dark dimension can also be understood as a semicircular dimension endowed with $S^1/\mathbb{Z}_2$ symmetry.\footnote{In the context of toroidal orientifold models compact dimensions are associated to $S^1/\mathbb{Z}_2$ symmetry which is needed for obtaining a 4D chiral model ($S^1$ does not lead to chirality).}

After the end of 5D inflation, the radion is stabilized in
a local (metastable) vacuum by the Casimir energy of 5D fields. More
concretely, the 
effective 4D potential of the radion $R$ after the end of inflation takes the form  
\begin{equation}
  V(R) = \frac{2 \pi \ \Lambda_5 \ r^2}{R} +\left(\frac{r}{R} \right)^2 \ T_4  + V_C (R) \,,
\label{V}
\end{equation}
where $\Lambda_5$ is the 5D cosmological constant, $r \equiv \langle
R \rangle$, $T_4$ is the total brane tension, and
\begin{equation}
V_C (R) = \pi R\left({r\over R}\right)^2 \sum_i (-1)^{s_i} \ N_i \
\rho(R,m) 
\label{VC}
\end{equation}
corresponds to the Casimir energy, and where the sum goes over all 5D states
 in the spectrum, $N_i$ is the number of degrees of freedom
 of the $i$-th particle, $s_i = 0 \, (1)$ for bosons (fermions),  and the
 factor $(r/R)^2$ comes from the rescaling of the 4D metric in the
 transformation to the Einstein frame~\cite{Anchordoqui:2023etp}. The
Casimir energy density for a particle of mass $m$ is given by 
\begin{equation}
\rho(R,m)=-\sum_{n=1}^\infty {2m^5\over (2\pi)^{5/2}} {K_{5/2}(2\pi Rmn)\over (2\pi Rmn)^{5/2}}\,,
\end{equation}
where $K_{5/2}(x)$ is the Bessel
function~\cite{Arkani-Hamed:2007ryu}.\footnote{The Casimir energy per
  unit of volume is the same for Dirichlet and Neumann boundary conditions~\cite{Barone:2003rn}.} 

The Casimir contribution to the potential falls off exponentially at
large $R$ compared to the particle wavelength. Considering
 only the first two terms in (\ref{V}) it is straightforward to see that the
 potential develops a maximum at
\begin{equation}
 R_{\rm max} = - T_4/(\pi
 \Lambda_5) \,,
\end{equation}
requiring a negative tension $T_4$. 

Corrections to the vacuum energy due to Casimir forces are expected to become important in the
deep infrared region. Indeed,  as $R$ decreases different particle thresholds open up, 
\begin{equation}
V_C (R) =  \sum_i \frac{\pi r^2}{32
   \pi^7 R^6} \ (N_F - N_B) \ \Theta (R_i -R) \,,
 \label{xxx}
\end{equation} 
where $m_i =
 R_i^{-1}$, $\Theta$ is a step function,  $N_F - N_B$ stands for the
 difference between the number of light fermionic and bosonic degrees
 of freedom~\cite{Anchordoqui:2023etp}. Eventually, the fermionic degrees of freedom overwhelm the bosonic contribution, giving rise to possible
minima, as long as $R_i < R_{\rm max}$.

\begin{figure}[htb!]
    \postscript{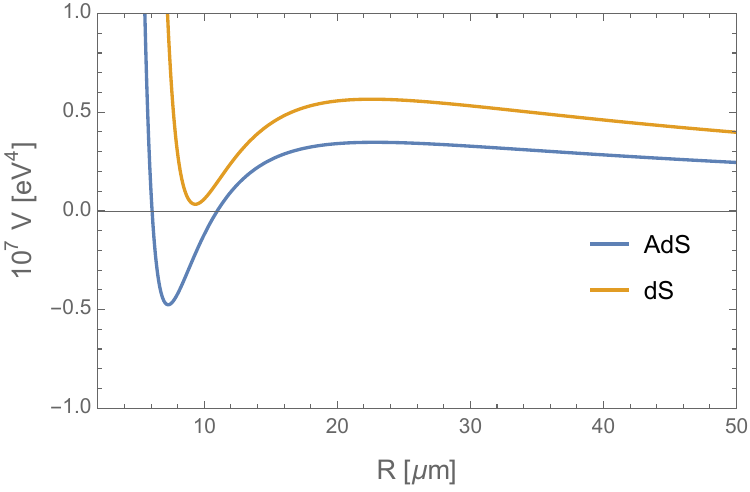}{0.9}
  \caption{The potential $V(R)$  for $\Lambda_5 =
    (22.6~{\rm meV})^5$ and $|T_4| = (24.2~{\rm meV})^4$ considering $N_F-N_B = 3$
    (AdS) and $N_F-N_B = 7$ (dS). 
   \label{fig:1}}
\end{figure}

To get further insights into this idea we envision a next-to-minimal hidden
sector to characterize the deep infrared region. It consists of: the 5D
graviton (contributing with $N_B = 5$), a massive gauge boson
($N_B=4$), and three massive bulk fermions (each contributing with $N_F = 4$). These fermions
could be identified with sterile neutrinos furnished by Dirac bulk
masses, which weaken the bounds on the size of the fifth dimension from neutrino oscillation data~\cite{Anchordoqui:2023wkm}. Altogether, this gives
a difference $N_F - N_B = 3$. Now, we would further assume that the
gauge boson is unstable and decays into two neutrinos. The 5D gauge
coupling has dimension of mass$^{-1/2}$, which we can write as
$g/M_*^{1/2}$ where $M_*$ is the 5D gravity scale (of order the string
scale) and $g$ a dimensionless parameter. The total decay
width is then given by
\begin{equation}
  \Gamma \sim \frac{g^2}{(2 \pi)^4} \frac{m^2}{M_*} \,,
\end{equation}
where $m$ is the mass of the gauge
boson. Taking $m \sim 100~{\rm meV}$,  $g \sim 10^{-4.5}$, and
$M_*\sim 10^9~{\rm GeV}$ we obtain $\Gamma
\sim 5 \times 10^{-42}~{\rm GeV}$, which implies the gauge bosons decay at $z \sim z_c$. If this
were the case, then for $z \lesssim z_c$ the difference in (\ref{xxx})  would
become $N_F - N_B = 7$. In Fig.~\ref{fig:1} we show an example of the
AdS $\to$ dS transition produced by the dark sector described above.

Now, the AdS $\to$ dS transition shown in Fig.~\ref{fig:1} slightly deviates from the model
analyzed in~\cite{Akarsu:2023mfb}, because the fields characterizing the deep
infrared region of the dark sector contribute to the effective number
of relativistic neutrino-like
species $N_{\rm eff}$~\cite{Steigman:1977kc}. Using conservation of entropy, fully
thermalized relics with $g_*$ degrees of freedom contribute
\begin{equation}
  \Delta N_{\rm eff} = g_* \left(\frac{43}{ 4g_s}\right)^{4/3} \left
    \{ \begin{array}{ll} 4/7 & {\rm for  \ bosons}\\ 1/2 & {\rm for \
                                                           fermions} \end{array}
               \right.                                        \,,
\end{equation}
where $g_s$ denotes the effective 
degrees of freedom for the entropy of the other thermalized
relativistic species that are present when they decouple~\cite{Anchordoqui:2019amx}. The 5D
graviton has 5 helicities, but the spin-1 helicities do not have zero
modes, because we assume the compactification has
$S^1/\mathbb{Z}_2$ symmetry and so the $\pm 1$ helicities are
projected out. The spin-0 is the
radion and
the spin-2 helicities form the massless (zero mode) graviton. This means
that for the 5D graviton, $g_*=3$. The $\pm 1$ helicities of the gauge
field are odd and do not have zero modes, only the two scalars are
even, so  $g_*=2$. The (bulk) left-handed neutrinos are odd, but the right-handed neutrinos are even and so each
counts as a Weyl neutrino, for a total $g_* =2 \times 3$. Assuming that the
dark sector decouples from the Standard Model sector before the electroweak phase
transition we have $g_s = 106.75$. This gives $\Delta N_{\rm eff} =
0.27$, consistent with the 95\%CL bound reported by the Planck Collaboration~\cite{Planck:2018vyg}. 

\begin{figure}[htb!]
    \postscript{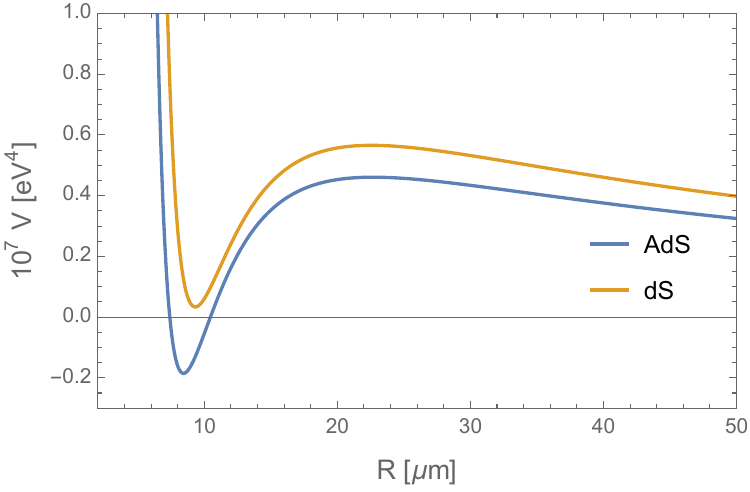}{0.9}
  \caption{The potential $V(R)$  for $\Lambda_5 =
    (22.6~{\rm meV})^5$ and $|T_4| = (24.2~{\rm meV})^4$ considering $N_F-N_B = 5$
    (AdS) and $N_F-N_B = 7$ (dS). 
   \label{fig:2}}
\end{figure}

We can also envision a more complex dark sector in which the gauge
field is replaced by a complex scalar with Yukawa coupling to the
neutrinos (that could provide them with mass). The spectrum has
addition of an extra complex scalar, such that before its decay  $N_F - N_B = 5$ and after decay $N_F - N_B = 7$. In
Fig.~\ref{fig:2} we show the corresponding AdS $\to$ dS transition for
this scenario. For this scenario, $\Delta N_{\rm eff} = 0.27$.

The next-to-minimal dark sectors that can accommodate the AdS $\to$ dS
crossover transition can
be nicely combined with the dark dimension proposal to elucidate the
radiative stability of the cosmological hierarchy $\Lambda/M_p^{4}
\sim 10^{-120}$~\cite{Montero:2022prj}. As can be seen in
Figs.~\ref{fig:1} and \ref{fig:2}, the connection is possible for a compactification radius $R_\perp^{\rm (
  AdS)} \sim
8~\mu{\rm m}$, which implies the Kaluza-Klein (KK) tower of the dark
dimension opens up at
$m_{\rm KK}^{\rm (AdS)} \sim 155~{\rm meV}$. After the vacuum
undergoes the AdS $\to$ dS transition the
compact space slightly enlarges to $R_\perp^{\rm (dS)} \sim 9~\mu{\rm
  m}$ and so $m_{\rm KK}^{\rm (dS)} \sim 138~{\rm meV}$. This in turn makes a shift of the species scale,
\begin{equation}
  M_* \sim (m_{\rm KK}^{(i)})^{1/3} M_p^{2/3} \,,
\end{equation}
 where $i =$ AdS, dS. Putting all this together, we have
this interesting observation that as the species scale goes down correspondingly the species entropy \cite{Cribiori:2023ffn}
 goes up, and we can even speculate that this has a deeper
 origin: since the entropy should increase we argue that {\it any AdS to dS
 transition in which the minimum is shifted towards smaller values of
 $R$ are rule out}. Note that the reduction in the species scale can
be also understood as a result of fixing the 4D $M_p$. Fixing instead
 the 5D scale $M_*$  amounts to increase $M_p$ and thus
 reduce the 4D strength of gravity since the radius becomes
 bigger. This alternative interpretation is constrained by
 observations, see e.g.~\cite{Alvey:2019ctk}. 

\begin{figure}[htb!]
    \postscript{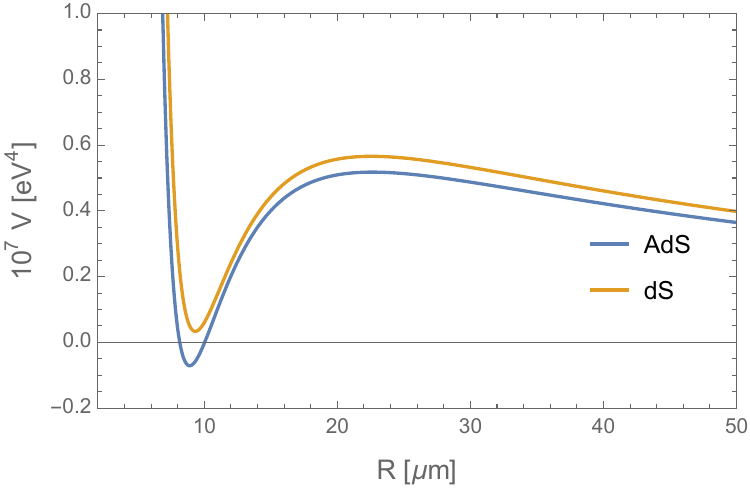}{0.9}
  \caption{The potential $V(R)$  for $\Lambda_5 =
    (22.6~{\rm meV})^5$ and $|T_4| = (24.2~{\rm meV})^4$ considering $N_F-N_B = 6$
    (AdS) and $N_F-N_B = 7$ (dS). 
   \label{fig:3}}
\end{figure}

Another possible course of
   action to accommodate an AdS $\to$ dS transition (with
   $N_F - N_B =6$ in the AdS phase and $N_F - N_B =7$ in the dS phase)
   is to assume that a real scalar
$\varphi$ has a potential with two local minima with very small difference in
vacuum energy and bigger curvature (mass) of the lower one. More
concretely, the potential is given by
\begin{equation}
V(\varphi) = \xi \varphi - \frac{1}{2} m^2 \varphi^2 +
\frac{\zeta}{3!} \varphi^3 +
\frac{\lambda}{4!} \varphi^4
\,,
\end{equation}
where the parameters $\xi$, $m^2$, $\zeta$, and $\lambda$ are real and
positive. The potential has two minima ($\varphi_{\rm fv}$ and
$\varphi_{\rm tv}$, the false and true vacuum, respectively) whose
free-energy difference is non-zero. We fix the model parameters by requiring
\begin{equation}
m_{\varphi_{\rm fv}} \sim \sqrt{V_{\varphi\varphi} (\varphi_{\rm fv})}
  \sim 100~{\rm meV} \,,
\end{equation}  
\begin{equation}
m_{\varphi_{\rm tv}} \sim \sqrt{V_{\varphi\varphi} (\varphi_{\rm tv})}
  \sim 500~{\rm meV} \, ,
\end{equation}
and
\begin{equation}
\Delta V \equiv V(\varphi_{\rm fv}) -
V(\varphi_{\rm tv}) \ll (\Lambda_5)^{4/5} \, ,
\end{equation}
where $V_\varphi = \partial V/\partial \varphi$. The fourth degree of
freedom is adjusted by requiring the
height of the barrier to accommodate a decay rate of the false vacuum
to be
\begin{equation}
\Gamma = 2 \ \hslash \ {\rm Im}\left\{\lim_{T\to \infty} \frac{1}{T} \
  \ln \left[ Z(T) \right] \right\}
\sim 5 \times 10^{-42}~{\rm GeV} \, ,
\end{equation}
where $Z (T) = \int [dx] \ e^{- S_E/\hslash}$ is the path integral, $[dx]$ is a functional measure, and $S_E$ the
Euclidean action~\cite{Coleman:1980aw,Hawking:1981fz}. Finally, we take $\Lambda_5 \sim (22.6~{\rm meV})^5$ and $|T_4| = (24.2~{\rm
  meV})^4$ to obtain the vacuum energy densities of the AdS and dS phases shown
in Fig.~\ref{fig:3}. For this scenario, $\Delta N_{\rm eff} = 0.25$.

At this stage, it is constructive to connect with contrasting and
complementary perspectives to comment on two caveats of our analysis:
{\it (i)}~although the cosmological models are motivated by swampland
constraints and string constructions, they have no concrete UV
completion; {\it (ii)}~another aspect of this analysis which may
seem discrepant at first blush is the fact that the (one loop) Casimir energy
is a zero-$T$ effect. However, in the example of the real scalar
(which not unexpectedly is at zero
temperature) the argument to understand the
transition is
essentially the same than the one in finite temperature models because the number of light degrees of freedom
changes due to a different transition of the 5D scalar field. In other
words, the model avoids finite temperature requirements and relies on
an ordinary vacuum decay in five dimensions. This obviously implies that the AdS
vacuum is not a true vacuum. The vacuum in the radius modulus is
determined by the contribution to the Casimir potential of the number
of light degrees of freedom. This number changes discontinuously due
to an ordinary vacuum decay of a 5D scalar field which satisfies the
swampland AdS conjecture (for instance it can be dS to dS). This
change drives the AdS to dS transition in the radius modulus, which is therefore discontinuous as in first order transitions.

It is of interest to further investigate the impact of adding
extra-relativistic degrees of freedom into the $\Lambda_s {\rm CDM}$
model. In particular, a value of $\Delta N_{\rm eff} \neq 0$ would
accelerate the universe before the CMB epoch, and hence could modify
the critical redshift $z_c$ for the AdS $\to$ dS crossover
transition. A study along this line is presented in an 
accompanying paper~\cite{Anchordoqui:2024gfa} where we use the Boltzmann
solver \CLASS~\cite{Blas:2011rf,Lesgourgues:2011rh} in combination with \MP~\cite{Audren:2012wb,Brinckmann:2018cvx} (and the latest results
from the {\it Planck} mission~\cite{Planck:2019nip}, Pantheon+ SNe Ia~\cite{Brout:2022vxf}, BAO~\cite{Nunes:2020hzy,deCarvalho:2021azj}, and
KiDS-1000~\cite{KiDS:2020suj}) to perform a Monte Carlo
Markov Chain study of the $\Lambda_s {\rm CDM} + N_{\rm eff}$ model and determine 
regions of the parameter space that can simultaneously resolve the $H_0$ and $S_8$ tensions. \\

\section*{Acknowledgments}

We thank Niccol\`o Cribiori,  Eleonora Di Valentino, and Arthur Hebecker for valuable
discussions. The research of LAA is supported by the U.S. National
Science Foundation (NSF grant PHY-2112527). The work of D.L. is
supported by the Origins Excellence Cluster and by the
German-Israel-Project (DIP) on Holography and the Swampland.

\end{document}